\documentstyle[12pt,epsfig]{article}

\catcode`\@=11
\def\@citex[#1]#2{%
\if@filesw \immediate \write \@auxout {\string \citation {#2}}\fi
\@tempcntb\m@ne \let\@h@ld\relax \def\@citea{}%
\@cite{%
  \@for \@citeb:=#2\do {%
    \@ifundefined {b@\@citeb}%
      {\@h@ld\@citea\@tempcntb\m@ne{\bf ?}%
      \@warning {Citation `\@citeb ' on page \thepage \space undefined}}%
      {\@tempcnta\@tempcntb \advance\@tempcnta\@ne%
      \@tempcntb\number\csname b@\@citeb \endcsname \relax%
      \ifnum\@tempcnta=\@tempcntb 
        \ifx\@h@ld\relax%
          \edef \@h@ld{\@citea\csname b@\@citeb\endcsname}%
        \else%
          \edef\@h@ld{\ifmmode{-}\else--\fi\csname b@\@citeb\endcsname}%
        \fi%
      \else
        \@h@ld\@citea\csname b@\@citeb \endcsname%
        \let\@h@ld\relax%
      \fi}%
    \def\@citea{,\penalty\@highpenalty\,}%
  }\@h@ld
}{#1}}

\def\@citeb#1#2{{[#1]\if@tempswa , #2\fi}}
\def\@citeu#1#2{{$^{#1}$\if@tempswa , #2\fi }}
\def\@citep#1#2{{#1\if@tempswa , #2\fi}}

\def\bcites{         
        \catcode`\@=11
        \let\@cite=\@citeb
        \catcode`\@=12
}

\def\upcites{         
        \catcode`\@=11
        \let\@cite=\@citeu
        \catcode`\@=12
}

\def\plaincites{      
        \catcode`\@=11
        \let\@cite=\@citep
        \catcode`\@=12
}

\newcount\hour
\newcount\minute
\newtoks\amorpm
\hour=\time\divide\hour by 60
\minute=\time{\multiply\hour by 60 \global\advance\minute by-\hour}
\edef\standardtime{{\ifnum\hour<12 \global\amorpm={am}%
        \else\global\amorpm={pm}\advance\hour by-12 \fi
        \ifnum\hour=0 \hour=12 \fi
        \number\hour:\ifnum\minute<10 0\fi\number\minute\the\amorpm}}
\edef\militarytime{\number\hour:\ifnum\minute<10 0\fi\number\minute}

\def\draftlabel#1{{\@bsphack\if@filesw {\let\thepage\relax
   \xdef\@gtempa{\write\@auxout{\string
      \newlabel{#1}{{\@currentlabel}{\thepage}}}}}\@gtempa
   \if@nobreak \ifvmode\nobreak\fi\fi\fi\@esphack}
        \gdef\@eqnlabel{#1}}
\def\@eqnlabel{}
\def\@vacuum{}
\def\marginnote#1{}
\def\draftmarginnote#1{\marginpar{\raggedright\scriptsize\tt#1}}
\def\draft{
        \pagestyle{plain}
        \overfullrule=2pt
        \oddsidemargin -.5truein
        \def\@oddhead{\sl \phantom{\today\quad\militarytime} \hfil
        \smash{\Large\sl DRAFT} \hfil \today\quad\militarytime}
        \let\@evenhead\@oddhead
        \let\label=\draftlabel
        \let\marginnote=\draftmarginnote
        \def\ps@empty{\let\@mkboth\@gobbletwo
        \def\@oddfoot{\hfil \smash{\Large\sl DRAFT} \hfil}
        \let\@evenfoot\@oddhead}
        \def\@eqnnum{(\theequation)\rlap{\kern\marginparsep\tt\@eqnlabel}%
        \global\let\@eqnlabel\@vacuum}  }

\def\blackfonts{
        \font\blackboard=msbm10 scaled\magstep1
        \font\blackboards=msbm8
        \font\blackboardss=msbm6
}

\def\nblack{            
        \def\ZZ{{Z \n{10} Z}}
        \def\NN{{N \n{14} N}}
        \def\CC{{C \n{11} C}}
        \def\RR{{R \n{11} R}}
        \def\QQ{{Q \n{12} Q}}
        \def\PP{{P \n{11} P}}
}

\def\prep{         
        \catcode`\@=11
        \input art10.sty
        \catcode`\@=12
        
        \let\small\null
        \def\blackfonts{
                \font\blackboard=msbm10
                \font\blackboards=msbm7
                \font\blackboardss=msbm5
        }
        \let\sl\it
        \twocolumn
        \sloppy
        \voffset=-2.54truecm
        \hoffset=-2.54truecm
        \flushbottom
        \parindent 1em
        \leftmargini 2em
        \leftmarginv .5em
        \leftmarginvi .5em
        \marginparwidth 48pt
        \marginparsep 10pt
        \setlength{\columnsep}{2truecm}
        \setlength{\textwidth}{25.4truecm}
        \setlength{\textheight}{17truecm}
        \baselineskip=16pt
        \oddsidemargin .18truein
        \evensidemargin .17truein
}

\def\eqalign#1{\null\,\vcenter{\openup\jot\m@th
  \ialign{\strut\hfil$\displaystyle{##}$&$\displaystyle{{}##}$\hfil
      \crcr#1\crcr}}\,}
\def\eqalignno#1{\displ@y \tabskip\centering
  \halign to\displaywidth{\hfil$\@lign\displaystyle{##}$\tabskip\z@skip
    &$\@lign\displaystyle{{}##}$\hfil\tabskip\centering
    &\llap{$\@lign##$}\tabskip\z@skip\crcr
    #1\crcr}}

\def\section{\@startsection {section}{1}{\z@}{3.ex plus 1ex minus
 .2ex}{2.ex plus .2ex}{\large\bf}}
\def\subsection{\@startsection{subsection}{2}{\z@}{2.75ex plus 1ex minus
 .2ex}{1.5ex plus .2ex}{\bf}}        

\def\appendix{{\newpage\section*{Appendix}}\let\appendix\section%
        {\setcounter{section}{0}
        \gdef\thesection{\Alph{section}}}\section}

\def\abstract{\if@twocolumn
\section*{Abstract}
\else 
\begin{center}
{\bf Abstract\vspace{-.5em}\vspace{0pt}}
\end{center}
\quotation
\fi}

\catcode`\@=12

\newcommand{\beq}{\begin{equation}}
\newcommand{\eeq}{\end{equation}}
\newcommand{\beqa}{\begin{eqnarray}}
\newcommand{\eeqa}{\end{eqnarray}}

\newcommand{\Z}{{\bf Z}}

\newcommand{\R}{{\bf R}}

\newcommand{\e}{{\rm e}}

\newcommand{\dd}{{\rm d}}

\def\noj#1,#2,{{\bf #1} (19#2)\ }
\def\jou#1,#2,#3,{{\sl #1\/ }{\bf #2} (19#3)\ }
\def\ann#1,#2,{{\sl Ann.\ Physics\/ }{\bf #1} (19#2)\ }
\def\cmp#1,#2,{{\sl Comm.\ Math.\ Phys.\/ }{\bf #1} (19#2)\ }
\def\ma#1,#2,{{\sl Math.\ Ann.\/ }{\bf #1} (19#2)\ }
\def\ng#1,#2,{{\sl Nagoya.\ Math.\ J.\/ }{\bf #1} (19#2)\ }
\def\jd#1,#2,{{\sl J.\ Diff.\ Geom.\/ }{\bf #1} (19#2)\ }
\def\invm#1,#2,{{\sl Invent.\ Math.\/ }{\bf #1} (19#2)\ }
\def\cq#1,#2,{{\sl Class.\ Quantum Grav.\/ }{\bf #1} (19#2)\ }
\def\cqg#1,#2,{{\sl Class.\ Quantum Grav.\/ }{\bf #1} (19#2)\ }
\def\ijmp#1,#2,{{\sl Int.\ J.\ Mod.\ Phys.\/ }{\bf A#1} (19#2)\ }
\def\jmphy#1,#2,{{\sl J.\ Geom.\ Phys.\/ }{\bf #1} (19#2)\ }
\def\jams#1,#2,{{\sl J.\ Amer.\ Math.\ Soc.\/ }{\bf #1} (19#2)\ }
\def\grg#1,#2,{{\sl Gen.\ Rel.\ Grav.\/ }{\bf #1} (19#2)\ }
\def\mpl#1,#2,{{\sl Mod.\ Phys.\ Lett.\/ }{\bf A#1} (19#2)\ }
\def\nc#1,#2,{{\sl Nuovo Cim.\/ }{\bf #1} (19#2)\ }
\def\np#1,#2,{{\sl Nucl.\ Phys.\/ }{\bf B#1} (19#2)\ }
\def\pl#1,#2,{{\sl Phys.\ Lett.\/ }{\bf #1B} (19#2)\ }
\def\pla#1,#2,{{\sl Phys.\ Lett.\/ }{\bf #1A} (19#2)\ }
\def\pr#1,#2,{{\sl Phys.\ Rev.\/ }{\bf #1} (19#2)\ }
\def\prd#1,#2,{{\sl Phys.\ Rev.\/ }{\bf D#1} (19#2)\ }
\def\prl#1,#2,{{\sl Phys.\ Rev.\ Lett.\/ }{\bf #1} (19#2)\ }
\def\prp#1,#2,{{\sl Phys.\ Rept.\/ }{\bf #1C} (19#2)\ }
\def\ptp#1,#2,{{\sl Prog.\ Theor.\ Phys.\/ }{\bf #1} (19#2)\ }
\def\ptpsup#1,#2,{{\sl Prog.\ Theor.\ Phys.\/ Suppl.\/ }{\bf #1} (19#2)\ }
\def\rmp#1,#2,{{\sl Rev.\ Mod.\ Phys.\/ }{\bf #1} (19#2)\ }
\def\yadfiz#1,#2,#3[#4,#5]{{\sl Yad.\ Fiz.\/ }{\bf #1} (19#2) #3%
\ [{\sl Sov.\ J.\ Nucl.\ Phys.\/ }{\bf #4} (19#2) #5]}
\def\zh#1,#2,#3[#4,#5]{{\sl Zh.\ Exp.\ Theor.\ Fiz.\/ }{\bf #1} (19#2) #3%
\ [{\sl Sov.\ Phys.\ JETP\/ }{\bf #4} (19#2) #5]}

\def\beq{\begin{equation}}
\def\eeq{\end{equation}}
\def\beqar{\begin{eqnarray}}
\def\eeqar{\end{eqnarray}}

\newcommand{\be}{\begin{equation}}
\newcommand{\ee}{\end{equation}}
\newcommand{\bea}{\begin{eqnarray}}
\newcommand{\eea}{\end{eqnarray}}

\def\nfrac#1#2{{\displaystyle{\vphantom1\smash{\lower.5ex\hbox{\small$#1$}}%
        \over\vphantom1\smash{\raise.25ex\hbox{\small$#2$}}}}}

\def\d#1{{}_{#1}}

\def\n#1{\mskip-#1mu}

\def\to{\rightarrow}

\def\lae{\mathrel{\mathop{\smash{\lower .5 ex \hbox{$\stackrel<\sim$}}}}}
\def\lae{\mathrel{\mathop{\smash{\lower .5 ex \hbox{$\stackrel>\sim$}}}}}

\def\l:{\mathopen{:}\,}
\def\r:{\,\mathclose{:}}

\catcode`\@=11
\def\theequation{\arabic{equation}}
\catcode`\@=12

\nblack
\bcites

\nblack

\catcode`\@=11
\def\theequation{\thesection.\arabic{equation}}
\@addtoreset{equation}{section}
\@addtoreset{footnote}{section}
\@addtoreset{footnote}{subsection}
\catcode`\@=12

\def\d{{\rm d}}

\def\ddbar{D$p$-anti-D$p$}

\newcommand{\beqn}{\begin{equation}}
\newcommand{\eeqn}{\end{equation}}
\newcommand{\beqnarray}{\begin{eqnarray}}
\newcommand{\eeqnarray}{\end{eqnarray}}
\newcommand {\bear} [1] {\begin {array} {#1}}
\newcommand {\ear} {\end {array}}

\newcommand {\beqarn} {\begin{eqnarray*}}
\newcommand {\eeqarn} {\end{eqnarray*}}

\begin{document}

\begin{titlepage}

\begin{center}
\hfill CALT-68-2263 \\
\hfill CITUSC/00-010\\
\hfill HUTP-99/A063\\
\hfill KIAS-P00002\\
\hfill  hep-th/0002223

\vskip 0.5 cm
{\Large \bf Confinement on the Brane}
\vskip 1 cm 
{Oren Bergman$^{*}$\footnote{Electronic mail: bergman@theory.caltech.edu}, 
Kentaro Hori$^{\dagger}$\footnote{Electronic mail: hori@infeld.harvard.edu}
and Piljin Yi$^{\#}$\footnote{Electronic mail: piljin@kias.re.kr}}\\
\vskip 0.5cm

$^{*}${\sl California Institute of Technology,
Pasadena, CA 91125, USA\\
and\\
CIT/USC Center for Theoretical Physics, \\
Univ. of Southern California,
Los Angeles CA}

\vskip 5mm
$^\dagger${\sl Jefferson Physical Laboratory, Harvard University,
Cambridge, MA 02138}

\vskip 5mm
$^\#$ {\sl School of Physics, Korea Institute for Advanced Study\\
207-43 Cheongryangri-Dong, Dongdaemun-Gu, Seoul 130-012, Korea}

\end{center}

\vskip 0.5 cm
\small
\begin{abstract}
A non-perturbative confinement mechanism has been proposed to explain the 
fate of the unbroken gauge group on the world-volume of annihilating 
D-brane-anti-D-brane pairs. In this paper, we examine this phenomenon 
closely from several different perspectives. Existence of the confinement 
mechanism is most easily seen by noticing that the fundamental string emerges 
as the confined electric flux string at the end of the annihilation process. 
After reviewing the confinement proposal in general, this is shown explicitly 
in the D2-anti-D2 case in the M-theory limit. Finally, we address the crucial 
issue of whether and how confinement occurs in the weakly coupled limit of 
string theory. 
\end{abstract}

\end{titlepage}

\normalsize

\section{Introduction}

D-branes and anti-D-branes are believed to undergo an annihilation process
analogous to that of ordinary particles and anti-particles. 
However, unlike ordinary particles, 
D-branes carry local degrees of freedom associated with open strings.
An obvious question is what happens to these degrees of freedom when the 
branes and anti-branes annihilate. Given that we anticipate a supersymmetric 
vacuum without the branes, it is clear that all such
open string modes must be removed from the spectrum of 
physical states, leaving behind a pure closed string theory.\footnote{Up 
to lower dimensional branes as by-products of the annihilation process.}
One would like to understand this process from the open string
point of view.

In the weak string coupling regime, and
at energies well below the string scale ($\alpha'^{-1/2}$), D-brane
dynamics can be approximated by concentrating on the lowest lying modes 
of the open strings, and ignoring the coupling to bulk closed strings.
The result is a world-volume field theory \cite{Leigh,Witten},
which includes, among other
fields, world-volume gauge fields. 
In the brane-anti-brane system the world-volume theory contains
one gauge field on the brane and another on the anti-brane, as well 
as a  complex tachyonic scalar from the brane-anti-brane string.
Both gauge fields must be removed from the spectrum in the annihilation 
process. The question we would like to ask is
whether and how this process can be effectively 
described at the level of the world-volume gauge theory.

We can get some insight into this question by considering the
possible by-products of such an annihilation. It is well-known
that a non-vanishing world-volume gauge field strength serves as a source
for various space-time fields \cite{DCS,strominger}. 
Once the D-brane world-volume disappears, these fields must be supported by
remnant space-time objects.  For example, magnetic flux
on a D$p$-brane is a source for the R-R $(p-1)$-form field, and
therefore \cite{Polchinski} induces
a D$(p-2)$-brane charge. Charge conservation therefore
implies that after the annihilation one is left with precisely such a D-brane.
This requires, however, that there exists a {\em localized}
magnetic vortex solution in the world-volume theory, which suggests
that the annihilation process incorporates a Higgs mechanism.\footnote{
This argument is due to Kimyeong Lee \cite{kimyeong}.}

This is precisely what happens. The tachyon plays the role of
the Higgs field, and the Higgs mechanism proceeds via tachyon
condensation \cite{sentachyon,nonbps}.
This process induces a mass for a linear combination of the gauge 
fields, thereby removing it from the low-energy spectrum.
However, the other combination, under which the tachyon is neutral,
remains unbroken and appears to stay massless \cite{Sred,K}. This is
the puzzle of the unbroken $U(1)$.

To understand what happens to the unbroken $U(1)$,  we consider
this time electric flux on the D$p$-brane (on a circle), which serves as a 
source for the NS-NS 2-form field. The flux induces a (wound) fundamental 
string charge \cite{Witten}, so
in this case one must be left with a fundamental string
after the annihilation. From the world-volume perspective, this
requires electric flux to be {\em confined} to a thin tube, where
the confinement scale is set by the tension of the fundamental string. 
This suggests that the world-volume description of the annihilation 
process should also incorporate a confinement mechanism.
In fact, one of the authors has proposed in \cite{piljin}
that confinement indeed occurs and removes the unbroken $U(1)$ from the
low energy dynamics. For $p\geq 3$, it was argued that confinement is 
driven by a dual Higgs mechanism, involving the condensation of magnetically 
charged tachyonic states, which are realized as D$(p-2)$-branes suspended 
between the D$p$-brane and the anti-D$p$-brane. However, since this mechanism 
is non-perturbative it is difficult to establish rigorously at weak string 
coupling. Furthermore, it does not seem to be applicable directly to 
the cases with $p<3$. This paper will, in part, address these difficulties.

More recently it has been proposed by A. Sen that the removal of the
additional gauge sector might be understood at tree level in string 
perturbation theory \cite{sen}. Sen has shown, under the crucial assumption
that the minimum of the perturbative tachyon potential precisely cancels 
the tension of the brane, that the kinetic term for the unbroken gauge
field vanishes when the tachyon attains its vacuum expectation values.
This in turn implies that charged states are removed from the spectrum, 
which also suggests a form of confinement, albeit one that is
completely perturbative in origin.

We shall argue in section 4 that this observation  is actually
incomplete by itself to resolve the puzzle of the unbroken $U(1)$. 
For instance, 
it cannot explain why the electric flux condenses into fundamental 
strings. Instead, it will actually help us to realize the non-perturbative 
world-volume confinement picture of Ref.~\cite{piljin} much more concretely. 
Specifically, we will use the result of \cite{sen} for the D1-anti-D1 and 
D2-anti-D2 systems, and derive the tension of the confined electric flux 
tube, thereby confirming its identification with the fundamental string. 
In the process, we propose an alternate scenario for the tachyon potential, 
where the supersymmetric vacuum is restored only when both perturbative and 
non-perturbative tachyonic directions are turned on.

It is the purpose of this paper to offer additional arguments
in favor of the proposal of \cite{piljin}, and study its implication
in the weakly coupled limit of string theory. In particular we will 
clarify the relationship between this non-perturbative confinement 
mechanism and Sen's recent observations on the 
perturbative effective action of
unstable D-branes. In section 2 we review the proposed 
confinement mechanism in the brane-anti-brane system,
and propose the analogous mechanism for the unstable
D-branes of Type II  string theory. In section 3, we 
describe the D2-anti-D2
case in purely geometric terms using M-theory, and demonstrate explicitly 
the production of fundamental strings as confined electric flux.
We then apply a similar geometrical approach to other systems, and
derive additional by-products of brane-anti-brane annihilation.
In section 4 we will come back to the all important question of how
the (non-perturbative) confinement occurs in the weak coupling limit, 
and how Sen's observation fits into this phenomenon.

\section{The Confinement Mechanism}

The confinement mechanism in question involves non-perturbative objects 
like open D$(p-2)$ branes; it is rather difficult to probe
in the perturbative regime of string theory.
Nevertheless, there is ample evidence that such a
mechanism should exist in the brane-anti-brane
annihilation process, not the least of which is that
it would solve the puzzle of the unbroken $U(1)$.
One of more compelling pieces of evidence is the charge conservation 
argument outlined in the introduction. 
The qualitative dynamics of confinement can be most easily
understood as a Higgs mechanism of an antisymmetric tensor field 
which is dual to the confined gauge field \cite{piljin}. 
We will start by reviewing the original proposal in Section 2.1,
which holds for  D$p$-anti-D$p$
pairs for $p \ge 3$. ($p\le 2$ cases are special because of the
low dimensional nature, and needs a different approach. See Section
3.1 and Section 4.) This will be followed by a discussion of
confinement in other cases, such as
in multiple brane-anti-brane pairs and unstable D-branes.

\subsection{Brane-anti-Brane}

The \ddbar\ system includes in its
world-volume two gauge fields $A$ and $A'$
on the brane and on the anti-brane,
respectively, and a complex tachyonic scalar
$T$ from strings ending on both. To determine the charge of $T$, recall that
the world-volume
interaction which assigns charge to the
endpoint of a string is given by
\begin{equation}
\int *F^{(2)}\wedge B^{(2)}\;,
\label{fstring_end}
\end{equation}
where $B^{(2)}$ is the (pullback of the) NS-NS 2-form field.
The endpoints of open strings will thus carry electric charge with respect
to the world-volume gauge fields $A$ on D$p$ and $A'$ on anti-D$p$. 
The relative sign of  two charges is a matter of
convention, which we will choose by saying that the open string
is neutral under the symmetric combination $A_+ = A+A'$. With this
convention, the open string carries a unit
electric charge with respect to $A_- = A-A'$. 

The low-energy world-volume theory is
therefore an Abelian Higgs model with a gauge group $U(1)\times U(1)$
broken spontaneously to the diagonal subgroup $U(1)_+$.
As the tachyon is expected to condense to a value of the order
of the string scale, this generates an ${\cal O}(\alpha'^{-1/2})$
mass term for $A_-$, thereby removing it from the low-energy spectrum.
A by-product of this Higgs effect is a magnetic vortex.
In an Abelian Higgs model, the topological soliton arises from 
the winding number of the scalar expectation values at
infinity, which induces quantized magnetic flux. 
This magnetic vortex can be shown to carry quantized D$(p-2)$-brane 
charges, owing to the topological term \cite{DCS},
\begin{equation}
\int F^{(2)}\wedge C^{(p-1)} \;,
\label{dstring_end}
\end{equation}
where $C^{(p-1)}$ is the (pullback of the) R-R $(p-1)$-form field.

For $p\geq 3$, the world-volume theory also contains non-perturbative 
magnetically charged  objects, corresponding to D$(p-2)$-branes
which are suspended between the brane and the anti-brane
\cite{strominger}.
Their charges are determined by the topological term in (\ref{dstring_end}),
which tells us that the boundaries are magnetic monopoles.
Since the two couplings in (\ref{fstring_end}) and (\ref{dstring_end}) 
have opposite
parities under orientation reversal of the world-volume,
the non-perturbative states carry  magnetic charge with respect to 
$A_+$ and are neutral under $A_-$ \cite{piljin}. 

With the exception of the $p=3$ case, we do not know much more
about these magnetic states, since they correspond to extended
objects in the world-volume. In particular, their tension should
in principle be determined by the ground state energy of the
suspended D$(p-2)$-brane, which we do not know in general.
For $p=3$ the magnetic state is a particle, which corresponds
to a suspended D-string. Here too, we cannot in general compute
the mass of the state. At large string coupling ($g_s\gg 1$) however,
the quantization of the D-string is the same as the quantization of
the fundamental string for $g_s\ll 1$. Thus for $g_s\gg 1$ the
ground state of this D-string corresponds to a magnetically charged
tachyon $\widetilde{T}$. Since this tachyon is charged only under
$A_+$, its condensation leads to a dual Higgs mechanism,
in which the dual gauge field $\widetilde{A}_+$ becomes massive.
In the original variables, this translates into confinement. 
The ``solitonic'' by-product of confinement is a thin electric flux 
string, which, due to the coupling in (\ref{fstring_end}), carries
the fundamental string charge.

For $p>3$ the situation is more complicated. The question is how to describe
higher-dimensional analogs of the dual Higgs mechanism, whereby an extended 
magnetic state becomes tachyonic ({\em i.e.} negative tension-squared)
and condenses. Generalization of the Higgs mechanism to higher rank 
anti-symmetric tensor fields is straightforward  
\cite{piljin,rey}, however, the difficulty lies in describing the magnetically 
charged tachyonic objects. 

On the other hand, it turns out that for the special case of $p=4$ one 
can give an indirect argument that there should be such a generalized
Higgs mechanism for anti-symmetric tensor fields on branes \cite{piljin}. 
The D4-anti-D4 system in Type IIA string theory
corresponds to an M5-anti-M5 system wrapped on
the circle in the eleventh dimension of M-theory.
The ordinary Higgs mechanism in the Type IIA system must therefore
lift to an analogous mechanism for the self-dual and
anti-self-dual two-forms, which live on the M5-brane and
anti-M5-brane, respectively. 

This also implies, in particular, that M5-anti-M5 
annihilation can produce M2-branes as solitonic by-products.
This follows from the fact that, in the Type IIA picture,
a transverse M2-brane is a D2-brane, 
which is realized as a world-volume vortex in the D4-anti-D4 system.
On the other hand, a wrapped M2-brane  
corresponds to a fundamental string in Type IIA, which
appears in the D4-anti-D4 system as confined electric flux of $A_+$. From the
perspective of D4-anti-D4, this confinement proceeds via a dual Higgs
mechanism for $A_+$, as anticipated.
This establishes that for the D4-anti-D4 system the Higgs and confinement 
phenomena are described by a single, Higgs-like, effect on an 
M5-anti-M5 pair.
So, at least for this system, we see that both mechanisms do occur.

This example touches upon the other crucial question of whether
the (perturbative) Higgs mechanism and the (non-perturbative) dual Higgs 
mechanism occur simultaneously. Naively, one would expect that when
one sector is weakly coupled the other is necessarily strongly coupled,
due to the duality transformation. This would imply that
we can discuss only one of the two sectors reliably, which may
cast some doubt on the resolution of the puzzle of the unbroken $U(1)$ 
by confinement. The 
above example of D4-anti-D4 alleviates some of this doubt, given that the
two phenomena are actually the same effect in M-theory,
but it cannot be generalized to other cases. Thankfully, however,
it turns out that this problem is naturally resolved. The two mechanisms,
for a very interesting reason, can be simultaneously described 
in weak-coupling descriptions. 
We will come back to this crucial question in Section 4.

\subsection{Multiple Brane-anti-Brane}

For a system of $N$ \ddbar\ pairs the gauge 
group is $U(N)\times U(N)$, and the ``electric'' tachyon transforms
in the bi-fundamental representation, {\em i.e.}
$({\bf N},\overline{\bf N})\oplus(\overline{\bf N},{\bf N}) $.
The candidate for the ``magnetic'' tachyon corresponds to an open
D$(p-2)$-brane, and transforms in the bi-fundamental representation
of the dual gauge group. 
As the former condenses, the gauge group is broken to the diagonal 
subgroup $U(N)_+$, and we must somehow
explain how the magnetic tachyon confines the remaining non-Abelian 
gauge sector.

The argument for a single pair is not applicable here. 
The main obstacle is the question of how the 
magnetic tachyon transforms under the dual unbroken gauge group.
In fact, it is not clear how to dualize non-Abelian gauge fields
to begin with. One way to get around this problem is to break the original
non-Abelian group
to its Cartan subgroup $U(1)^{2N}$ by turning on the adjoint scalar
fields corresponding to separating the different \ddbar\ pairs.
In this Coulomb branch, the physics of brane-anti-brane annihilation
is then a simple generalization of the single \ddbar\ case.
Namely, $N$ of the $U(1)$ factors are broken by the Higgs mechanism,
and the others are confined.
The $N$ confined $U(1)$ gauge fields belong to the diagonal
subgroup $U(N)_+$, so the dual Higgs mechanism confines
the gauge charges associated with the $N$ mutually commuting 
generators of $U(N)_+$. 
At the origin of the Coulomb branch the $U(N)_+$ symmetry is restored,
and the Cartan generators are no longer singled out.
This suggests, but does not prove, that the entire non-Abelian gauge
sector will be confined.

\subsection{Unstable D-branes}

We would now like to extend the confinement picture to the unstable 
D-branes of Type II string theory \cite{senunstable,horava}.
These are D$p$-branes where $p$ has the ``wrong'' values, 
namely $p$ is odd in Type IIA and even in Type IIB. 
They correspond to boundary states 
which have components in the NS-NS sector but not in the R-R sector.
Consequently, the open string spectrum
consists of an unprojected NS sector
and an unprojected R sector, and includes
a real neutral tachyon as well
as a massless $U(1)$ gauge
field \cite{senunstable,horava}.\footnote{For $N$ coincident unstable
D-branes the gauge group is $U(N)$,
and the tachyon transforms in the
adjoint representation.}
Here one encounters an apparent puzzle in trying
to apply the confinement mechanism for the unbroken gauge group,
which in this case is just the $U(1)$.
Since the D$(p-2)$-brane, like the D$p$-brane,
does not carry R-R charge,
it does not appear to give rise to any (magnetic)
world-volume charge on the 
D$p$-brane.

The resolution lies in the realization of the unstable D$p$-brane
of Type IIA(B) as a projection of the \ddbar\ system in Type IIB(A)
by the discrete symmetry generated by $(-1)^{F_L}$,
where $F_L$ denotes the left-moving part
of the space-time fermion number \cite{senunstable}. 
This follows directly from the action of $(-1)^{F_L}$ on the 
Chan-Paton factors of the open strings in the \ddbar\ system. 
In particular, the lowest lying (bosonic) states are given by
\begin{equation}
\left[
\begin{array}{ll}
 A & T \\
 T^* & A'
\end{array}
\right]\;,
\end{equation}
out of which only the combinations $A_+=A+A'$ and $T+T^*$ survive
the projection. Since the tachyon is neutral under $A_+$, these
are precisely the lowest-lying degrees of freedom
of the unstable D-brane.

We now show that there is indeed a $(p-3)$-dimensional object in the 
world-volume that is magnetically charged under the $U(1)$ gauge group.
Recall that before the projection, the relevant magnetic state
was produced by a (BPS) D$(p-2)$-brane suspended
between the D$p$-brane
and the anti-D$p$-brane. 
Since all R-R fields are odd under the operator $(-1)^{F_L}$,
D-branes are mapped to anti-D-branes, and vice-versa.
In particular, the D$(p-2)$-brane is mapped to an 
anti-D$(p-2)$-brane, but at the same time
the D$p$-brane and anti-D$p$-brane are exchanged.
As a result, the magnetic state actually survives the projection.
As shown in the previous section, this state is magnetically
charged under
$A_+$ and neutral under $A_-$. It is therefore magnetically
charged under
the gauge field on the unstable D-brane.
As in the brane-anti-brane system, it can therefore lead  to confinement of
the gauge field by its condensation.

\section{Geometric Confinement}

In the previous section we tried to
understand the confinement mechanism
from the world-volume field theory viewpoint.
On the other hand, one crucial
aspect of confinement, namely the existence of
a confined electric flux string and its
identification with the fundamental string,
can also be seen from the space-time
perspective. As mentioned in the introduction,
one merely needs to invoke
charge conservation to see that electric flux
must condense into  
fundamental strings. In particular,
there are cases where this process
can be seen explicitly from purely geometric considerations,
which we propose to call
``geometric confinement.''  These are cases where electric flux can be 
translated into a winding of branes along a compact circle, such as the
M-theory circle along $X^{11}$. 

\subsection{D2-anti-D2}

Confinement through the dual Higgs effect does not 
apply directly to the system of D2-anti-D2. One may imagine that a related, 
Polyakov type mechanism works for this 2+1-dimensional system. This will
be addressed in Section 4. Independently of this, however, 
confinement on D2-anti-D2 can be seen in another way, it turns out.
Let us first recall that
a D2-brane is equivalent to an M2-brane that is transverse
to the 11-th dimensional circle \cite{M}.
The gauge field on D2 is then realized as the dual of the periodic 
scalar field $\eta\equiv X^{11}/R_{11}$ on the 2+1-dimensional world-volume
of the membrane,
\begin{equation}
\partial_i A_j-\partial_j A_i=\epsilon_{ijk}\partial_k\eta,
\end{equation}
where the derivatives are with respect to the three world-volume coordinates
$x^i$. With this in mind, consider an electric flux 
configuration along the direction of $x^1$, 
\begin{equation}
\partial_0 A_1-\partial_1 A_0=f(x^2)=\partial_2\eta .
\end{equation}
Using this relationship to solve for $\eta$, we find that the 
M2-brane is actually winding around the circle along $X^{11}$,
\begin{equation}
X^{11}(x^2)=R_{11}\int^{x^2}_{-\infty}
f(s)ds .
\end{equation}
Now suppose that we have a D2-anti-D2 pair, with a unit of 
flux on D2 (so that the corresponding
M2 brane winds around the compact circle once) and no flux on anti-D2.
We assume that both membranes are asymptotically flat
($X^1= x^1$ and $X^2=x^2$). 
The situation is illustrated in Figure 1.

\begin{figure}[htb]
\begin{center}
\epsfxsize=4in\leavevmode\epsfbox{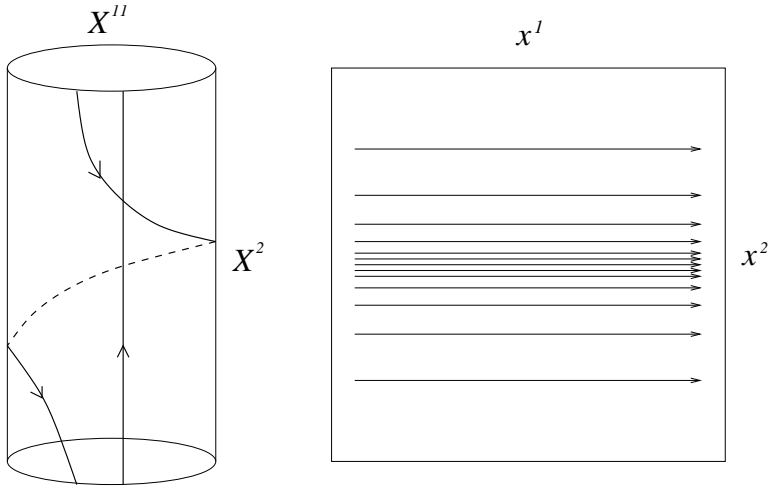}
\end{center}
\caption{An M2-brane winding around $X^{11}$ once and an anti-M2-brane
with no winding (left). On the D2-brane, the winding translates into a unit
of electric flux (right).}
\label{fig1}
\end{figure}

Now let us ask what happens to this flux
once the pair annihilates.
The annihilation process dissipates
the tension of the branes: the more complete
the process, the lower  the
energy. On the other hand, the winding cannot
be removed in this process,
and at the end one finds a remnant which is an M2-brane
tightly wrapped around
the circular $X^{11}$ direction.
Scanning along  the $x^2$ direction, we
find that before the annihilation
the flux $\partial_2\eta$ is distributed
over some finite width along $x^2$, while
after the annihilation
the flux is practically localized at some definite $x^2$. From the world-volume
perspective, one finds a confined electric flux string. From the space-time
viewpoint, this is nothing but the fundamental string.

\begin{figure}[htb]
\begin{center}
\epsfxsize=4in\leavevmode\epsfbox{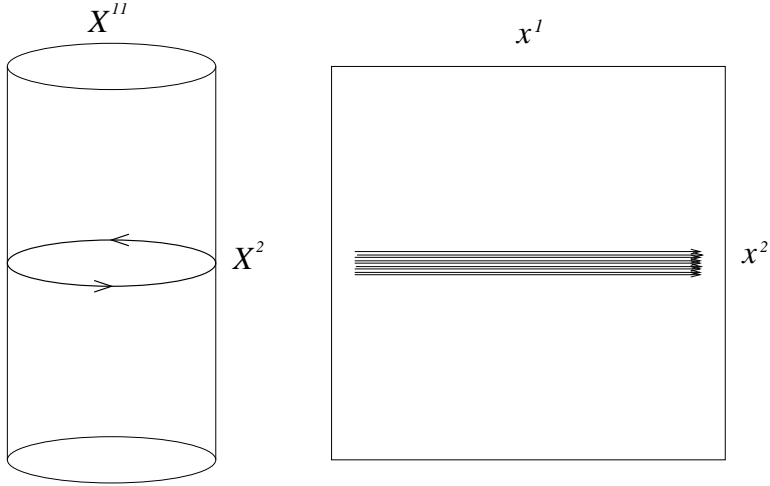}
\end{center}
\caption{Most of the M2-brane annihilates with the anti-M2-brane, 
but part of it remains and winds around $X^{11}$. The result is a single 
longitudinal M2-brane along $X^{11}$ (left).
Seen from the original transverse 
world-volume of the M2-brane (right),
this translates into a tightly confined
flux string.}
\label{fig2}
\end{figure}

We discovered that certain electric  flux is confined upon D2-anti-D2 
annihilation, and that the resulting confined string is the fundamental  
string of Type IIA theory. This shows that confinement of the world-volume 
gauge field indeed occurs as part of the annihilation process.

One may still ask if this indeed corresponds
to confinement of the correct
$U(1)$ gauge field. Recall that one combination
of $U(1)$'s should be actually
Higgsed. To see this clearly,
let us recall the fact that anti-branes are 
nothing but branes with opposite orientation.
This distinction shows up
in Hodge-dual operations one performs to convert the
periodic scalar to a
vector field. If we denote the periodic scalar and
its dual gauge field 
on the anti-M2-brane by $\eta'$ and $A'$, we find
\begin{equation}
\partial_i A'_j-\partial_j A'_i=
-\epsilon_{ijk}\partial_k\eta' .
\end{equation}
Note the sign on the right hand side. The net relative winding lies in
the linear combination $\eta-\eta'$, so the confined gauge field
is the linear combination $A_+=A+A'$. 

On the other hand,
from the decomposition of the 11-dimensional metric
to the 10-dimensional metric
plus dilaton and R-R 1-form gauge field ${\cal C}^{(1)}$,
we learn that 
the sum of the two Nambu-Goto actions for the two membranes
contains a term,
\begin{equation}
\int\:(dA-dA')\wedge {\cal C}^{(1)} .
\end{equation}
Thus, it is the magnetic flux of the difference of
the two gauge fields that
generates D0-brane charge \cite{sentachyon}. This is precisely the gauge
field identified by Sen
as being Higgsed by the perturbative tachyon,
whose magnetic vortex generates
the D0 charge. The other linear combination
$A+A'$ is left intact by 
the perturbative sector of Type II superstrings, and its electric flux 
corresponds to the net winding in figures 1 and 2. This is precisely the
world-volume gauge field that was argued to be confined. 

In effect, we established confinement for the  case of 
D2-anti-D2, at least when the M-theory description is appropriate.
Because the above confinement of the electric
flux can be seen from a geometrical viewpoint 
of space-time, we will call this phenomenon ``geometric confinement.''

\subsection{More Decay Channels of Brane-anti-Brane Pairs}

While we started with the D2-anti-D2 system (motivated 
by the proposal of Ref.~\cite{piljin}), 
the above mechanism easily generalizes to other cases.
Whenever one finds 
a $(p+1)$-dimensional brane-anti-brane pair transverse
to a compact circle, 
one can dualize the compact scalar $\xi$ associated
with the position of 
the $p$-brane to a $(p-1)$-form tensor field, 
\begin{equation}
\partial_i\xi=\frac{1}{(p-1)!}\;
\epsilon_i^{\:j_1\cdots j_{p-1}}{\cal A}_{j_1\cdots j_{p-1}}.
\end{equation}
Defining $\xi'$ and its dual ${\cal A}'$ 
analogously for the anti-$p$-brane,
${\cal A}+{\cal A}'$ will undergo a similar geometric confinement process.
The confined
``electric'' flux forms a domain-wall. From the space-time perspective,
the domain  wall is simply a $p$-brane of the same kind
wrapped along the compact
 direction. An interesting example is the case of M5-anti-M5-branes 
transverse to $X^{11}$, namely an NS5-anti-NS5-brane pair, where the 
resulting domain-walls are wrapped M5-branes, also known as D4-branes. 

Starting with a D$p$-anti-D$p$ pair that is transverse to a circle $X^9$,  
the confined ``electric'' flux of ${\cal A}+{\cal A}'$ 
corresponds to a D$p$-brane which is wrapped
along $X^9$. Actually, the compact ``scalar'' $\xi- \xi'$ associated with 
the relative position of the pair along $X^9$ is not really a scalar. In the 
T-dual picture, it is part of the gauge field, where the phenomenon is 
annihilation of a D$(p+1)$-anti-D$(p+1)$ pair through condensation of 
the open string tachyon. The T-dual of the wrapped D$p$-brane is a transverse 
D$(p-1)$-brane, which is realized as a magnetic vortex on the longitudinal 
D$(p+1)$-anti-D$(p+1)$ world-volume. Only when the compact direction 
$R_9$ is large compared to the string scale, we may consider $\xi$ as a 
scalar field. In that limit, the dual circle along ${\tilde X}^{9}$
has a small radius $1/R_9$, and a vortex on D$(p+1)$-anti-D$(p+1)$ 
world-volume is de-localized along this compact direction,
and effectively 
becomes a domain-wall configuration. Thus, one may regard geometric
confinement on a D$p$-anti-D$p$ pair as a special limit of Sen's
tachyon condensation, seen from a T-dual perspective.

Finally, we also believe that there is a mechanism which is complementary 
to geometric confinement. It is only one linear combination of 
${\cal A}+{\cal A}'$ that is confined by the latter. The other combination 
${\cal A}-{\cal A}'$ must acquire a mass-gap by some other means as well. 
In the case of D$p$-anti-D$p$ transverse to $X^9$,
this mechanism would be T-dual to confinement of the vector field already
proposed in Ref.~\cite{piljin}. Since the fundamental
string wrapped along $X^9$ transforms into a KK momentum mode along $X^9$,
the solitonic by-product one obtains is a Kaluza-Klein momentum mode. 
Analogously, we expect the possible 
solitonic by-products of the M5-anti-M5 system transverse to $X^{11}$ to
include a D0-brane.\footnote{ Ref.~\cite{lozano} identified the couplings 
between the space-time and the world-volume fields which are necessary 
for these additional decay modes of M5-anti-M5.}
(In another special case of M2-anti-M2 pairs transverse to compact $X^{11}$, 
the mechanism is again nothing but Sen's open string tachyon condensation, 
whose solitonic by-product is a D0-brane.)

This, together with the ordinary confinement
mechanism of Ref.~\cite{piljin}, suggests
to us an interesting possibility.
When a D-brane and an anti-D-brane annihilate
in the presence of compact circles, certain
perturbative closed string states could emerge.
The states that we have found are all BPS
(they carry either a winding number or
a KK momentum of fundamental strings)
and thus relatively easy to 
probe. Taking this one step further,
one may be persuaded that the entire closed string 
spectrum should be reproduced from dynamics of open strings.

\section{Confinement at Weak String Coupling}

Since the world-volume
gauge theory description of the D-brane dynamics is
valid only at weak string coupling $g_s\ll 1$,
it is clear that the issue of the
unbroken $U(1)$ should be addressed
in this regime.
The main question concerning the confinement mechanism
as a possible resolution is that it involves highly non-perturbative
objects in string theory with a huge tension $\sim 1/g_s$.
In particular, it is not clear that such objects give
the most important effect.
Similarly, ``geometric confinement''
is established only at strong Type IIA coupling,
where a semi-classical description of the membranes is valid.
A priori, confinement at strong coupling need not imply 
confinement at weak coupling.

Recently, in the study of the effective action of unstable D$p$-branes
in Type II string theory,
Sen proposed a new mechanism to explain the
disappearance of the
unbroken $U(1)$ purely in terms of tree-level string theory \cite{sen}.
Using the results of \cite{SW} on brane dynamics
in the background of a constant magnetic or $B$ field,
it was shown that
the bosonic part of the effective action
in terms of the $U(1)$ gauge field strength
$F=dA_+$ and the tachyon field $T$,
is given by
\beq
S=-{1\over g_s}\int \dd^{p+1}x \sqrt{-\det\,(g+F)}\,V(T) ,
\label{Sens}
\eeq
under the assumption that $F$ and $T$ are constant.
In the above expression $V(T)$ is the tachyon potential
and $g$ is the induced metric on the
world-volume, which is also assumed to be constant.
This can be extended to the system of D-brane and
anti-D-brane, where $F$ is the gauge field of the unbroken $U(1)$,
under the additional
assumption that the gauge field of the broken $U(1)$ vanishes
(we thank A. Sen for discussion on this point).
\footnote{We are using the 
normalization where the two endpoints of an open fundamental string 
carry unit charges with respect to the gauge fields $A$ and $A'$.
We also set $2\pi\alpha'=1$, so that the tension of the fundamental string
is $1$. Also, we take the convention that $V(T)/g_s$ at $T=0$ is the brane 
tension. Thus, $V(0)$ is
$\sqrt{2}(2\pi)^{{1-p\over 2}}$ for the unstable D$p$-brane and
$2(2\pi)^{{1-p\over 2}}$ for the D$p$-anti-D$p$-brane pair.}
Here we are assuming that the NS $B$-field is zero, but
it is useful to keep in mind that a non-zero
$B$-field would enter in (\ref{Sens}) as $F\to F+B$.
It has been argued that the tachyon potential $V(T)$ vanishes at its
bottom.
If this is true, 
the gauge kinetic term vanishes
at the bottom of the tachyon potential.
In \cite{sen} it was further
argued that this means that the $U(1)$ gauge field acts
as a Lagrange multiplier which removes all charged objects
from the spectrum.
Since this is a tree-level effect in string theory, it
appears to be a perfect resolution to the puzzle.
Is the non-perturbative
confinement mechanism of no relevance for the resolution?

\subsection{$p=1$ Case}

To examine the consequence of (\ref{Sens}) more carefully,
let us consider the simplest non-trivial example, $p=1$.
It may appear that the issue of the
unbroken $U(1)$ is absent in this case,
since $1+1$ dimensional gauge fields have no propagating
degrees of freedom.
However, there are actually topological degrees of freedom
when the theory is formulated on a compact space,
and these require consideration.

We consider a D1-anti-D1 pair, or an unstable D1-brane, wrapped on
a circle of radius $R$. The effective action is given by
(\ref{Sens}) with $p=1$, and the integral is over $\R\times S^1$.
We focus on the dynamics of the gauge field $A$, and fix all
other degrees of
freedom. In particular, we assume that the induced metric $g$ is
the diagonal matrix $g_{11}=-g_{00}=1$, $g_{01}=0$.
The system is then described by a single bosonic
gauge invariant variable $a$ of period $1/R$,
which parameterizes the holonomy on $S^1$
as $\int_{S^1}A=2\pi R\,a$.
The world-volume theory
is now reduced to the quantum mechanics
obtained from the Lagrangian
\beq
L=-{2\pi R\over \lambda}\,\sqrt{1-\dot{a}^2} ,
\eeq
where
\beq
{1\over \lambda}\,=\,{1\over g_s}\,V(T).
\eeq
This is identical to the Lagrangian for the free
relativistic particle of
mass $2\pi R/\lambda$.
Thus, the Hamiltonian is given by
\beq
H=\sqrt{\,p^2+(2\pi R/\lambda)^2},
\eeq
where $p$ is the momentum conjugate to $a$,
$p=\partial L/\partial\dot{a}
=(R/\lambda)\dot{a}/\sqrt{1-\dot{a}^2}$.
Since $a$ is a periodic variable of period $1/R$, the wave functions are
spanned by $\psi_n(a)=\e^{2\pi iRna}$, for integer $n$.
These are eigenstates of the momentum $p=-i\partial/\partial a$
with eigenvalues $2\pi R\,n$.
The energy of the state $\psi_n$ is given by
\beq
E_n=2\pi R\,\sqrt{n^2+{1\over \lambda^2}}.
\label{Em}
\eeq

The ground state is $\psi_0$ and has energy
\beq
E_0=R\times {1\over g_s}\,V(T).
\eeq
At the bottom of the tachyon potential, where
we assume $V(T)=0$,
the ground state has a vanishing energy, and can be identified
with the supersymmetric vacuum of the string theory.
Conversely, the identification of the ground state with
the supersymmetric string vacuum
requires the tachyon potential $V(T)$ to vanish
at its bottom.

Now consider the excited states $\psi_n$ with $n\ne 0$.
In the limit of vanishing $V(T)$, $\psi_n$ has a finite energy
\beq
E_n\,\longrightarrow 2\pi R|n| ,
\label{string}
\eeq
and remains in the spectrum.
Note that the momentum $p=\partial L/\partial
\dot{a}$ is also a source for
the Neveu-Schwarz B-field $B_{01}$ \cite{Witten}.
Thus $\psi_n$ must represent a state in string theory with
the fundamental string
wrapped $n$ times on $S^1$. 
Indeed, the energy (\ref{string}) 
is precisely the mass of an $n$-wound fundamental string.
So the electric flux ``tube'', which in this case
fills the $1+1$ dimensional world-volume, can truly be identified
with the fundamental string.

In the present discussion, we have implicitly assumed that
(\ref{Sens}) holds for arbitrary configurations of the field,
and ignored the probable higher derivative corrections. 
To be very precise this requires justification, even though
all the eigenstates have constant field strengths
in the present case.

\subsection{$p\geq 2$ Cases}

The above example explicitly shows
that Sen's result (\ref{Sens}) together with the assumption
that $V(T)=0$ at the bottom does not
really eliminate the unbroken $U(1)$ degrees of freedom,
but rather supports the idea of
confinement.
The key point was that
a massless relativistic particle can carry a non-zero energy. 
Even though the Lagrangian appears to vanish in the massless limit,
the Hamiltonian remains non-trivial.
We expect a similar and possibly more interesting story in the
$p>1$ cases.

  Actually, an important aspect was ignored in the
argument for the resolution of the puzzle as a direct consequence
of (\ref{Sens}).
The vanishing of the gauge kinetic term is equivalent to
the blow-up of the gauge coupling.
When the gauge coupling becomes large, one has to start
worrying about strong gauge dynamics, and it is usually
impossible to describe this in the original variable. In other words, the
description in terms of the gauge field does not really make sense
when the tachyon expectation value comes close to
the bottom of the potential.

When one description breaks down, one must go over to another, better,
description.
In the theory of a 1-form gauge potential in $(p+1)$-dimensions
with a large
coupling, the better description is in terms of the dual $(p-2)$-form
potential with inverse coupling.
The latter becomes better as the original coupling becomes
larger.
Now, it is clear that of utmost relevance is the object of
least mass or tension that is charged under the $(p-2)$-form potential.
For $p\geq 3$, it is precisely the
$(p-3)$-brane in the world-volume
coming from the stretched D$(p-2)$-brane.
It is therefore tempting  to consider
confinement of the unbroken $U(1)$
by the magnetic Higgs mechanism as the actual resolution of
the puzzle.
For $p=2$, this does not work since
the ``$(p-3)$-dimensional object'' does not exist
as a charged object.
However, Euclidean D0-branes can stretch between D2-branes,
and can modify the dynamics by an instanton effect.
We shall first consider this case in detail, postponing the $p\geq 3$
cases for later discussion.
Throughout the discussion
we assume that higher derivative corrections to (\ref{Sens})
can be ignored.

\subsection*{\it p~=~2}

It was shown by Polyakov in \cite{Polyakov}
that a $U(1)$ gauge theory in 2+1 dimensions
which includes monopole instantons
exhibits confinement by the instanton effect of the monopoles.
This is quite similar to the situation under consideration.
However, there are also
notable differences ---
we are considering the Born-Infeld action rather than
the standard Maxwell action,
and we do not know the size of the monopole-instanton
(``W-boson mass'')
or the value of the instanton action.
In particular, it appears hopeless to compute the
tension of the confined electric flux tube, even
if we could argue for confinement.
Nevertheless, under a certain assumption
on the instanton effect,
we can argue for confinement and
even compute the exact tension of the flux tube.

Let us first dualize the $U(1)$ gauge field ignoring the effect of
D-brane instantons.\footnote{See Ref.~\cite{arkady} for related
discussions on dualization of Born-Infeld action.}
We start with a system of a $U(1)$ gauge field $A$ and a one-form field
$\mit\Pi$ with the action
\beq
S^{\prime}=-{1\over\lambda}\int\dd^3x\sqrt{-g}
\sqrt{1+\lambda^2|{\mit\Pi}|^2}
+\int {\mit\Pi}\wedge F ,
\label{dualize}
\eeq
where $F=\dd A$ is the curvature of $A$, and
$|{\mit\Pi}|^2=g^{\mu\nu}{\mit\Pi}_{\mu}{\mit\Pi}_{\nu}$.
We first integrate over the one-form field $\mit\Pi$
in the stationary phase approximation.
The action $S^{\prime}$ is stationary at
\beq
{\mit\Pi}={1\over\lambda}{\ast F\over \sqrt{1-|F|^2}},
\label{PiF}
\eeq
where $\ast$ is the Hodge dual with respect to the metric $g_{\mu\nu}$,
and $|F|^2={1\over 2}
g^{\mu\nu}g^{\rho\kappa}F_{\mu\rho}F_{\nu\kappa}$.
Inserting this value into (\ref{dualize})
we obtain the action for $A$
\beq
S=-{1\over\lambda}\int\dd^3x\sqrt{-g}\sqrt{1-|F|^2}.
\label{action2}
\eeq
It is easy to see that this is equal to (\ref{Sens}) with $p=2$,
if $\lambda$ is given by
\beq
{1\over\lambda}={1\over g_s}V(T).
\eeq
Next, let us exchange the order of integration.
Integrating out the gauge field $A$,
we obtain a constraint that
$\mit\Pi$ is a closed one form with integral period on one-cycles.
In other words, $\mit\Pi$ can be written as
\beq
{\mit\Pi}=\dd\sigma,
\label{PiS}
\eeq
where $\sigma$ is a periodic variable of period $1$
(so that
$\e^{2\pi i\sigma}$ is a circle valued function).
Now the action in terms of $\sigma$ is
\beq
\widetilde{S}=-{1\over\lambda}\int\dd^3x
\sqrt{-g}\sqrt{1+\lambda^2|\dd\sigma|^2}.
\label{dual}
\eeq
If $\lambda$ were just the string coupling
this would of course be the same as 
the action of the wrapped membrane in M-theory on $\R^{10}\times S^1$
written in string units.
Let us consider the limit $\lambda\to\infty$ corresponding to
$V(T)\to 0$, but still with $g_s\ll 1$.
Then the dual action (\ref{dual}) has a finite limit
\beq
\widetilde{S}\to-\int\dd^3x\sqrt{-g}|\dd\sigma|
=-\int\dd^3x\sqrt{-g}
\sqrt{g^{\mu\nu}\partial_{\mu}\sigma\partial_{\nu}\sigma}.
\eeq
This is completely different from the membrane action
in $\R^{11}$, which one would obtain from the dual of the
D2-brane action by sending $g_s\to\infty$, keeping finite the
eleven-dimensional Planck length.
The classical energy density of a static configuration
is given by
\beq
{\cal E}
=\sqrt{g^{ik}\partial_i\sigma\,\partial_k\sigma} ,
\eeq
where $i,k$ are spatial coordinate indices.
In particular,
the ground state has zero energy.

Now let us turn on an electric flux $F_{01}$ in the $x^1$-direction
of the world-volume.
Comparison of (\ref{PiF}) and (\ref{PiS}) shows that this corresponds to
turning on $\partial_2\sigma$.
Namely,
\beq
\Delta\sigma:=\sigma(x^2=+\infty)-\sigma(x^2=-\infty)
=\int\limits_{-\infty}^{+\infty}
{1\over\lambda}{F_{01}\dd x^2\over \sqrt{1-|F|^2}}.
\label{DelS}
\eeq
This quantity can be identified as the charge of
fundamental strings stretched in the $x^1$-direction,
since the integrand of the right hand side
is the same as $\delta S/\delta F_{01}=\delta S/\delta B_{01}$.
The flux carries energy density along
the $x^1$-direction given by
\beq
T=\int\dd x^2|\partial_2\sigma|=|\Delta\sigma|,
\label{tension1}
\eeq
if $\sigma$ is a monotonic function.
Since $T=1$ for a unit charge $|\Delta\sigma|=1$,
it may appear that
the flux can really be identified as the fundamental string.
However, we note that  (\ref{tension1}) holds for {\it any} monotonic
function $\sigma$, which can spread out without changing the
tension.
Thus, the electric flux does not tend to squeeze into
a thin tube, and confinement does not appear to occur.
Furthermore, even if we force the flux to be confined in a tube,
the charge $\Delta\sigma$ is not quantized (unless the $x^2$-direction
is compact) and we cannot truly identify
the tube as the fundamental string.
\begin{figure}[htb]
\begin{center}
\epsfxsize=3.5in\leavevmode\epsfbox{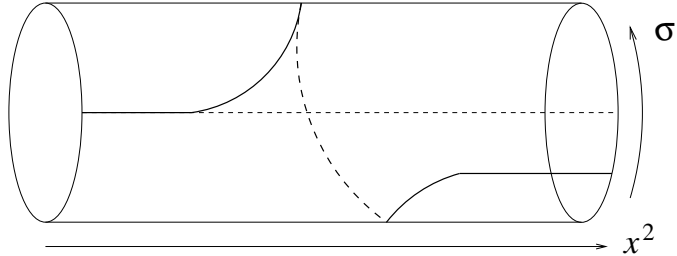}
\end{center}
\caption{{\it Any}
configuration attains minimum energy density (\ref{tension1})
as long as $\sigma(x^2)$ is a monotonic function.
Furthermore, the charge $\Delta\sigma$ is not quantized.}
\label{fig4}
\end{figure}

This can be cured by taking into account the effect of the
instantons.
For the D2-anti-D2 pair in Type IIA string theory,
a Euclidean trajectory of a D0-brane plays the role of
the instanton, whereas 
for the Type IIB unstable D2-brane,
the $(-1)^{F_L}$ projection of the Euclidean D0-anti-D0 pair
will do the job.
We recall that the instanton creates a point defect at $x$ in the
3-dimensional world-volume, from which a unit of magnetic flux
emanates. On a small sphere $S^2$ surrounding $x$
we have $\int_{S^2}F/2\pi=1$
(it would be $-1$ for an anti-instanton).
In the magnetic
description in terms of $\sigma$,
this corresponds to an insertion of the operator
$\e^{2\pi i\sigma(x)}$ in the path-integral
($\e^{-2\pi i\sigma(x)}$ for anti-instanton).
This can be seen as follows (this presentation is basically
from \cite{IAS}).
Insert the operators $\e^{2\pi i\sigma(x)}$
and $\e^{-2\pi i\sigma(y)}$ at distinct points $x\ne y$.
Because of (\ref{PiS}), in the original description (\ref{dualize})
this corresponds to the insertion of
\beq
\exp\left(2\pi i\int^x_y{\mit\Pi}\right) ,
\eeq
where the integral in the exponent is over any path starting at $y$
and ending at $x$.
This integral can be written as an integral over
the 3-dimensional world-volume $X$ as
\beq
\int\limits^x_y{\mit\Pi}=\int\limits_X{\mit\Pi}\wedge\omega ,
\eeq
where $\omega$ is a closed two form with $\delta$-function
support on the path, so that $\int_H\omega=1$
for any oriented hyperplane $H$ intersecting once with the path.
For example, $\omega$ has period $1$ and $-1$ on small 2-spheres
$S^2_x$ and $S^2_y$ surrounding $x$ and $y$, respectively.
Then the term $\int{\mit\Pi}\wedge F$ in (\ref{dualize}) is replaced by
$\int{\mit\Pi}\wedge(F+2\pi\omega)$, and the action after integration over
$\mit\Pi$ is given by (\ref{action2}) with $F$ replaced by
$F^{\prime}=F+2\pi\omega$.
Now, $F^{\prime}$ satisfies
\beq
\int\limits_{S^2_x}F^{\prime}/2\pi
=-\int\limits_{S^2_y}F^{\prime}/2\pi=1 ,
\eeq
and is therefore a curvature of a gauge field on $X-\{x,y\}$
which has a unit magnetic flux emanating from $x$ and going into $y$.
This shows the claim.

Summing over a gas of instantons and anti-instantons corresponds
to the insertion of
\beqa
\lefteqn{
\sum_{n_+,n_-}{1\over n_+!n_-!}\int
\prod_{i=1}^{n_+}\dd^3x_i\,\mu^3\,\e^{2\pi i\sigma(x_i)}
\prod_{j=1}^{n_-}\dd^3y_j\,\mu^3\,\e^{-2\pi i\sigma(y_j)}}\nonumber\\
&&
~~~~~~~~~~~~~~~~~~~~~~~~~~~~
=\exp\left(\,\mu^3
\int\dd^3 x\left(\e^{2\pi i\sigma(x)}+\e^{-2\pi i\sigma(x)}
\right)\,\right)
\eeqa
in the Euclidean path-integral, where $\mu$ is some mass parameter
which we assume to be non-zero.
The instanton effect therefore
generates a potential energy proportional to $\cos(2\pi\sigma)$,
and the new effective action in the limit $\lambda\to\infty$
becomes
\beq
\widetilde{S}_{\it eff}=
-\int\dd^3x\sqrt{-g}\left(
\sqrt{g^{\mu\nu}\partial_{\mu}\sigma\partial_{\nu}\sigma}
+M^3\cos(2\pi \sigma)\right),
\eeq
where $M$ is the $\lambda\to\infty$ limit of $\mu$, which is again
assumed to be non-zero.
The energy density of a static configuration is
then given by
\beq
{\cal E}_{\it eff}
=\sqrt{g^{ik}\partial_i\sigma\partial_k\sigma}
+M^3(\cos(2\pi\sigma)+1) ,
\eeq
where we have added a constant so that the system attains zero energy
in the ground state.
We do not know how to generate this energy shift at the moment,
but we take it for granted since the non-negativity of energy
is required on general grounds.

The formula (\ref{tension1}) for the energy density of the flux
is thus modified to
\beq
T=\int\dd x^2\,\Bigl(\,
|\partial_2\sigma|+\sqrt{g_{22}}M^3(\cos(2\pi\sigma)+1)\,\Bigr).
\label{tension2}
\eeq
In order for the tension to be finite, $\sigma$ must approach vacuum
values as $x^2\to\pm\infty$. The charge is therefore quantized,
\beq
\Delta\sigma=n\in\Z.
\eeq
Also, the potential term prevents
the flux from spreading, and confines it into a thin flux tube.
In the limit of a completely squeezed configuration,
the tension is equal to the flux,
\beq
T\to|n|.
\eeq
This is because the contribution from the potential vanishes in this limit.
In particular, the result is independent of the value of $M$, which
is difficult to estimate without a detailed knowledge of the
instanton. This is the magic of the limit $\lambda\to\infty$;
if $\sigma$ had a standard kinetic term $|\dd\sigma|^2$,
the stable configuration would have been determined by
a non-trivial balance between two effects
--- spreading by the kinetic term and squeezing by the potential term.
In our case, the spreading effect of the kinetic term vanishes
in the limit $\lambda\to\infty$.
\begin{figure}[htb]
\begin{center}
\epsfxsize=3.5in\leavevmode\epsfbox{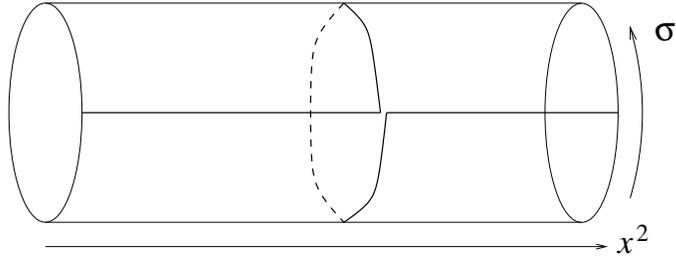}
\end{center}
\caption{By the instanton effect,
the charge $|\Delta\sigma|$ is quantized and
the energy density (\ref{tension2}) is minimized for
the completely squeezed configuration.
This density, or the tension of the resulting string,
is proportional to the charge.}
\label{fig5}
\end{figure}

To summarize,
we have seen that due to the effect of the D-brane instantons,
the electric flux is squeezed into a thin flux tube and is quantized.
The flux tube has the correct charge and tension to be identified with
the fundamental strings.

\subsection*{\it p~$\geq$~ 3}

\begin{figure}[htb]
\begin{center}
\epsfxsize=6in\leavevmode\epsfbox{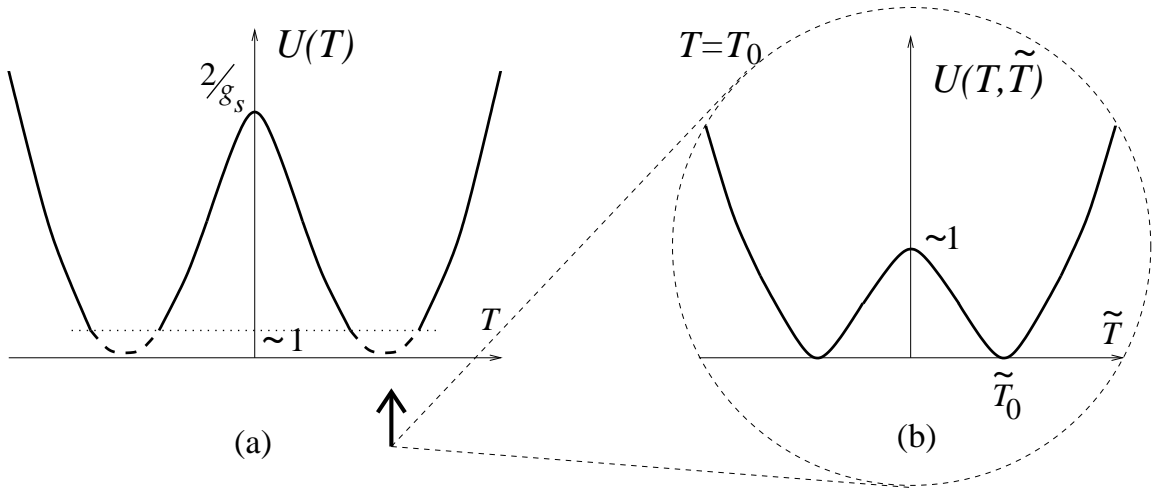}
\end{center}
\caption{The tachyon potential (a), and the tachyon-magnetic-tachyon
potential at the vacuum value of $T=T_0$ (b).}
\label{fig3}
\end{figure}

We finally discuss the $p\geq 3$ cases.
We do not attempt to make a quantitative analysis here.
However, we provide a possible picture of the physics 
in the weak string coupling regime $g_s\ll 1$ (See Figure \ref{fig3}).
To be specific, we consider a D$p$-anti-D$p$ system,
but the generalization to unstable D$p$-branes is obvious.
(We ignore the factors of $(2\pi)^{{1-p\over 2}}$ in this discussion.)

\begin{itemize}
\item
The system at $T=0$ has energy $2/g_s$.
Here both $U(1)$ gauge groups are unbroken,
and the corresponding gauge fields $A_-$ and $A_+$
are weakly coupled.
In particular the Lagrangian
for $U(1)_+$ is given by (\ref{Sens}) with
$V(T)=2$.
This point is unstable and the tachyon $T$ tends to
acquire non-zero values.

\item
Once the tachyon condenses $T\ne 0$,
$U(1)_-$ is broken by the Higgs mechanism,
and its gauge boson becomes massive.
$U(1)_+$ remains unbroken.
The potential energy
$U(T)=V(T)/g_s$
decreases as $T$ rolls down toward its vacuum value $T_0$.
But as long as $U(T)$ is much larger than 1, the
$U(1)_+$ gauge coupling is small,
and one can still use the gauge theory description of (\ref{Sens}).
Of course the tachyon continues to roll down to
smaller values of $U(T)$.

\item
When $U(T)$ comes close to 1,
the description in terms of the gauge
field $A_+$ is no longer appropriate, and 
one should use the dual magnetic description.
Standard electric-magnetic duality
turns the gauge field $A_+$ to a $(p-2)$-form gauge potential
$\widetilde{A}_+$.

\item
A $(p-3)$-dimensional object charged under
$\widetilde{A}_+$
is created by the stretched D$(p-2)$-brane,
and we denote the corresponding ``field'' by
$\widetilde{T}$.
Thus the system in this region is described by the fields
$(T,A_-)$ and $(\widetilde{T},\widetilde{A}_+)$,
which are possibly coupled by a potential $U(T,\widetilde{T})$.
Since we have dualized at $U(T)\sim 1$, we expect the potential energy at
$\widetilde{T}=0$ to be of order 1.
The dynamics is hard to analyze, but
we assume that $\widetilde{T}$ is tachyonic 
at $\widetilde{T}=0$ in constant-$T$ hyperplanes,
and tends to acquire non-zero expectation values
 (See Figure \ref{fig3} (b)).

\item
Once the magnetic tachyon condenses $\widetilde{T}\ne 0$,
the magnetic $U(1)_+$ is broken by the Higgs mechanism,
and the gauge boson $\widetilde{A}_+$
becomes massive. In other words, the original $U(1)_+$ is
confined.
We assume that the potential $U(T,\widetilde{T})$
vanishes at its bottom ($(T_0,\widetilde{T}_0)$ in Figure
\ref{fig3} (b)),
and has positive second derivatives there.
The vacuum configuration is then indistinguishable from
the supersymmetric vacuum of Type II string theory.

\end{itemize}

We recall that
a vortex configuration of the tachyon $T$
with the gauge field $A_-$ is
identified with the BPS D$(p-2)$-brane \cite{sentachyon}.
This is not altered in the new picture.
One must ensure however that both $T$ and $\widetilde{T}$
attain their vacuum expectation values
away from the core of the vortex.
The contribution to the energy
from the $(\widetilde{T},\widetilde{A}_+)$ sector, even if exists,
is of order 1 and is negligible compared to the contribution
$\sim 1/g_s$ from the $(T,A_-)$ sector.
Thus, the vortex can have the
correct tension $1/g_s$ to be identified as the BPS D$(p-2)$-brane.

There is another topological defect which 
comes purely from the $(\widetilde{T},\widetilde{A}_+)$ sector,
with $T$ at its vacuum value everywhere in the world-volume.
This is a topologically non-trivial configuration such that
$\widetilde{A}_+$
has a quantized period over a $(p-2)$-dimensional sphere
surrounding a one-dimensional object in the world-volume.
This one-dimensional object
can be identified as the confined electric flux tube
of $U(1)_+$ via electric-magnetic duality.
This in turn is identified with the fundamental
string, as follows from the coupling of $F=\d A_+$
to the NS $B$-field. Indeed, since the typical energy scale of the
$(\widetilde{T},\widetilde{A}_+)$ system is of order 1,
once the open string tachyon roles down to its vacuum value $T_0$,
the topological defect has tension of order 1
(the point is that the topological defect requires $\widetilde{T}$
to deviate from the vacuum but $T$ can remain at $T_0$).
This is the correct value of the tension for the fundamental string.

\subsection{The Fate of Massless Scalars}

So far, we have focussed on the world-volume gauge fields.
However, for $p<9$, the issue of the massless gauge boson
for the unbroken $U(1)$ is actually only a part of the problem;
there are also $(9-p)$ massless scalars, which represent
the center of mass motion of the D$p$-anti-D$p$ pair or
the unstable D$p$-brane.
\footnote{In the case of D$p$-anti-D$p$ pair, the relative motion is
frozen by tachyon condensation, which gives
mass to the corresponding scalars.}
Thus, in order to really solve the problem,
we must clarify the fate of these massless scalars.
We will not attempt to fully
solve this problem. Instead, we examine the situation
in the simplest case of $p=0$, and briefly comment on the
case of $p\geq 1$.

Unlike for gauge bosons, the puzzle of the massless
scalars exists already for $p=0$.
To be specific, let us consider a D0-anti-D0 pair
in Type IIA string theory on $\R^{10}$.
After tachyon condensation,
the nine scalars representing the relative motion become massive,
while
the other nine scalars $\phi^i$ ($i=1,\ldots,9$) representing
the center of mass motion remain massless.
The Lagrangian for the massless fields is
\beq
L=-{1\over \lambda}\sqrt{1-\sum_{i=1}^9(\dot{\phi}^i)^2}.
\eeq
This describes a relativistic particle in $\R^{10}$ of
mass $1/\lambda=V(T)/g_s$. Its mass therefore vanishes
at the bottom of the tachyon potential,
where it is assumed that $V(T)=0$.
In other words, {\it the fate of the center of mass scalar fields
in the D0-anti-D0 pair is to produce a massless particle
in $\R^{10}$}.
It is natural to interpret the latter as a massless particle in
the closed string spectrum. 
In fact, if the D0-anti-D0 pair annihilates and
no open string mode remains,
this is the only possible interpretation.

The fate of the center of mass scalars is less clear for
$p\geq 1$.
However, they cannot be ignored altogether simply because
the Lagrangian vanishes at the bottom of the
tachyon potential.
To see this, let us consider the situation
where an electric flux of the gauge theory sector is turned on
and is confined into a thin tube.
The action of the system can be factorized as
\beq
S=-\int_{\R\times W}\dd^{p+1}x\,\sqrt{-\det g}\,
{1\over\lambda}\sqrt{\det(1+g^{-1}F)},
\eeq
where $W$ is the spatial part of
the $p+1$ dimensional world-volume.
Confinement of electric flux in the gauge sector
means that one can replace the factor
${1\over\lambda}\sqrt{\det(1+g^{-1}F)}$
by a delta function on $W$ supported along the one-dimensional
flux string
$C\subset W$.
The action can then be expressed as
\beq
S=-\int_{\R\times C}\dd^2\sigma\,\sqrt{-\det \gamma},
\eeq
where $\gamma$ is the metric induced on
the flux string, which is determined by the massless
scalar fields (restricted on $\R\times C$).
Thus, the massless scalar fields restricted to the
flux string
represent the motion of the flux string in directions
transverse to the world-volume of the D$p$-anti-D$p$ system.
These ``new'' degrees of freedom\footnote{It is new compared to
the motion of the string within the brane $W$. The latter should emerge
as Goldstone bosons associated with translation along $W$.}
are actually required in order to identify the flux string as
the fundamental string, since the fundamental string can move
in the full eight transverse dimensions.
Without this, the string would have been trapped in
a plane $W$ of co-dimension $(9-p)$.

While this example shows that the scalar degrees of freedom play an
important role, it stops short of explaining fully how the
$(p+1)$-dimensional 
scalar fields really reduce to 2-dimensional ones. Their action
vanishes outside
the string core, but as we have seen earlier, a vanishing action does
not automatically imply disappearance of the associated degrees of freedom.

\section{Conclusions}

We started with Sen's simple observation that brane-anti-brane annihilation
should lead to the supersymmetric vacuum of closed string theory, and tried to 
understand how (some of) the
open string degrees of freedom are removed from the 
low energy spectrum. The basic mechanisms that 
lift the massless gauge sectors of the lowest lying open string
modes have been identified as the perturbative Higgs effect 
\cite{sentachyon} and non-perturbative confinement \cite{piljin}. 
We have shown that the two effects are in fact linked:
the former process forces us, via Sen's effective action, to 
describe the unbroken gauge sector
using the dual non-perturbative degrees of freedom,
which naturally allow us to describe confinement of the 
unbroken $U(1)$ as a weakly coupled dual Higgs mechanism.
The combined effect is to lift all gauge sectors in the lowest lying
open string modes. A careful consideration along similar lines might 
reveal how the remaining massless scalar fields
are lifted as well.

More generally, we believe that confinement is
one of the
central ingredients in converting open string degrees of freedom to closed
string ones. To see this, imagine that
we have semi-classical open strings suspended between a pair of D-branes.
Consider two such strings of opposite orientations, which are
well separated.
On the world-volume of each D-brane,
a unit of electric flux emanates from the end 
of one string and converges on that of the other.
Introducing an anti-D-brane
to annihilate against one of the D-branes, we should find that no string
remains ending on the annihilated D-brane. How is this accomplished via
local interactions on the branes? The obvious answer is that the electric 
flux gets confined, and becomes a segment of fundamental string that
connects the two ends of the semi-classical open strings. We can
also consider an analogous process where
a semi-classical open string with both ends
on a D-brane is converted to a semi-classical
closed string, by annihilation of the D-brane against another.
This picture is quite suggestive.

In a sense, one of the most surprising aspects 
of brane-anti-brane annihilation 
is that the process can be described, at least partly,
from the world-volume perspective.
To understand the annihilation process more completely, one 
possible approach would be to solve the corresponding open string field 
theory \cite{sen,szb}. At the moment, it is not clear to what extent
such a program can be carried out, especially in the face of the 
non-perturbative processes we encountered.

We have found the importance of the result (\ref{Sens})
of \cite{sen} in the world-volume field theory approach.
To be precise, this result is exact only
for a constant field strength and a constant induced metric,
and higher derivative corrections are probable.
We have assumed that such corrections can be ignored in our argument for
confinement,
but of course this requires a considerable justification.
It is hoped that our result provides a strong motivation
to clarify and estimate the validity of (\ref{Sens})
using various methods, including open string field theory.

Other immediate problems include
generalizing our observations to other cases, 
such as D5-anti-D5 annihilation in Type I theory. One can
imagine that an open D3-brane
ending on the D5's would play a role\footnote{Absence of
a closed D3 in Type I theory does not exclude the possibility
of an open D3 ending
on D5's, in much the same way that the
absence of a closed string does not imply
the absence of open strings.}, but a
further difficulty arises from the fact that
the would-be-confined  gauge sector is of $Sp(1)$. We hardly understand what
it means to have a dual Higgs mechanism for $Sp(1)$. Another interesting 
question is whether there is analog of K-theory \cite{K}
for the confined gauge sector. It would be classified by ``K-theory'' of 
geometrical objects associated with the antisymmetric 
tensor field (possibly ``gerbes'') dual to the gauge field.

\section*{Acknowledgment}

We would like to thank C.-S. Chu,
S. Hirano, N. Itzhaki, K. Lee, J. Maldacena,
S-J. Rey, A. Sen, E. Silverstein, C. Vafa and B. Zwiebach,
for useful discussions. We would also like to thank the Aspen Center for
Physics where part of this work was carried out.
OB is supported in part by the DOE under grant no.
DE-FG03-92-ER 40701.
KH is supported in part by NSF-DMS 9709694.


\begin{thebibliography}{99}

\small
\parskip=0pt plus 2pt

\bibitem{Leigh} R.~G.~Leigh,
``Dirac-Born-Infeld action from Dirichlet sigma model'',
Mod.\ Phys.\ Lett.\  {\bf A4}, 2767 (1989).

\bibitem{Witten}
E.~Witten,
``Bound states of strings and p-branes'',
Nucl.\ Phys.\ {\bf B460} (1996) 335,
hep-th/9510135.


\bibitem{DCS}
M. Li,
``Boundary States of D-Branes and Dy-Strings'',
Nucl. Phys. {\bf B460} (1996) 351, hep-th/9510161; 
M. Green, J.A. Harvey and
G. Moore,
`I-brane inflow and anomalous couplings on D-branes'',
Class. Quant. Grav. {\bf 14} (1997) 47, hep-th/9605033.

\bibitem{strominger}
A. Strominger,
``Open p-Branes'',  
Phys.Lett. {bf B383} (1996) 44,
hep-th/9512059.

\bibitem{Polchinski} J.~Polchinski,
``Dirichlet-Branes and Ramond-Ramond Charges'',
Phys.\ Rev.\ Lett.\  {\bf 75} (1995) 4724, hep-th/9510017.

\bibitem{kimyeong}
K. Lee, Private communication

\bibitem{sentachyon}
A.~Sen,
``Tachyon condensation on the brane anti-brane system'',
JHEP {\bf 08} (1998) 012, hep-th/9805170;

\bibitem{nonbps}
A.~Sen, 
``SO(32) spinors of type I and other solitons on brane-anti-brane pair'',
JHEP {\bf 9809} (1998) 023, hep-th/9808141; 
``Stable non-BPS bound states of BPS D-branes'', 
JHEP {\bf 9808} (1998) 010, hep-th/9805019;
``Stable Non-BPS states in string theory'', 
JHEP {\bf 9806} (1998) 007, hep-th/9803194.


\bibitem{Sred}
M.~Srednicki,
``IIB or not IIB'',
JHEP {\bf 9808} (1998) 005, hep-th/9807138.

\bibitem{K}
E.~Witten, 
``D-Branes and K-Theory'', 
JHEP {\bf 9812} (1998) 019,
hep-th/9810188.

\bibitem{piljin}
P.~Yi,
``Membranes from five-branes and fundamental strings from Dp branes'',
Nucl.\ Phys.\ {\bf B550} (1999) 214,
hep-th/9901159.


\bibitem{sen}
A.~Sen,
``Supersymmetric world-volume action for non-BPS D-brane''",
JHEP {\bf 9910} (1999) 008, hep-th/9909062;
"Universality of the tachyon potential", hep-th/9911116.

\bibitem{szb} A.~Sen and B.~Zwiebach,
``Tachyon condensation in string field theory'',
hep-th/9912249; N.~Berkovits, 
``The tachyon potential in open Neveu-Schwarz string field theory'',
hep-th/0001084; N.~Berkovits, A.~Sen, and B.~Zwiebach, 
``Tachyon condensation in superstring field theory'',
hep-th/0002211.


\bibitem{rey}
S. Rey, 
``The Higgs Mechanism for Kalb-Ramond Gauge Field'',
Phys. Rev. {\bf D40} (1989) 3396.


\bibitem{senunstable} A.~Sen,
``BPS D-branes on non-supersymmetric cycles'',
JHEP {\bf 9812}, 021 (1998)
[hep-th/9812031];
``Non-BPS states and branes in string theory'',
hep-th/9904207.


\bibitem{horava}
P. Ho\v rava, ``Type IIA D-Branes, K-Theory, and Matrix Theory'',
Adv. Theor. Math. Phys. {\bf 2} (1999) 1373, hep-th/9812135.

\bibitem{M}
M.J. Duff, P.S. Howe, T. Inami, K.S. Stelle,
``Superstrings in D = 10 from supermembranes in D = 11'',
Phys. Lett. {\bf B191} (1987) 70;

P.~K.~Townsend, 
``The Eleven-Dimensional Supermembrane Revisited,''
Phys.\ Lett.\  {\bf B350} (1995) 184, hep-th/9501068; 
``D-branes from M-branes,''
Phys.\ Lett.\  {\bf B373} (1996) 68, hep-th/9512062.

\bibitem{lozano}
L. Houart and Y. Lozano, ``Type II Branes from Brane-Antibrane in M-theory'',
hep-th/9910266.

\bibitem{SW} N.~Seiberg and E.~Witten,
``String theory and noncommutative geometry'',
JHEP {\bf 9909}, 032 (1999), hep-th/9908142.

\bibitem{Polyakov}
A.M.~Polyakov,
``Quark confinement and topology of gauge groups'',
Nucl.\ Phys.\ {\bf B120} (1977) 429.

\bibitem{arkady}
A.~A.~Tseytlin, 
``Self-duality of Born-Infeld action and Dirichlet 
3-brane of type IIB superstring theory'',
Nucl.\ Phys.\  {\bf B469}, 51 (1996), hep-th/9602064.

\bibitem{IAS} E. Witten, Lecture at IAS, Princeton, 1997,
http://www.math.ias.edu/QFT/index.html.

\end{thebibliography}
\end{document}